\documentclass[12pt]{JHEP3}

\usepackage{amsmath}

\newcommand{\be}[1]{ \begin{equation}\label{#1} }
\newcommand{\ee}{\end{equation}}
\newcommand{\ben}[1]{\begin{eqnarray}\label{#1} }
\newcommand{\een}{\end{eqnarray}}

\newcommand{\p}{\partial}
\newcommand{\refb}[1]{(\ref{#1})}
\newcommand{\OO}{{\cal O}}
\newcommand{\FF}{{\cal F}}

\newcommand{\DD}{{\cal D}}

\newcommand{\hh}{{\mathfrak h}}
\newcommand{\WW}{{\cal W}}
\newcommand{\AAA}{{\cal A}}
\newcommand{\II}{{\cal I}}

\title{Holographic Renormalisation for the Spin-3 Theory and the (A)dS$_3$/CFT$_2$ correspondence}

\author{Shailesh Lal$^{a}$\footnote{shailesh DOT lal AT icts DOT res DOT in} and Bindusar Sahoo$^{b}$\footnote{bsahoo AT nikhef DOT nl}
\\
$^a$International Center for Theoretical Sciences -- TIFR, \\
$\;$TIFR Centre Building, Indian Institute of Science,\\
$\;$Bangalore 560012,\\ 
$\;$India. \\
$^b$NIKHEF theory group, Science Park 105,\\
$\;$1098 XG Amsterdam,\\
$\;$The Netherlands.\\
}

\abstract{We compute the two-point correlation functions for the spin-3 theory in three dimensional (Anti-) de Sitter spacetimes by using holographic renormalisation. For the AdS case, we find results consistent with the general requirements of two-dimensional conformal invariance. In the de Sitter case, we find similar results. We discuss consistency requirements on the three point functions $\left<TWW\right>$ for our results to be compatible with the asymptotic symmetry algebra for AdS case and with the de-Sitter central charge found in hep-th/0106113  by analyzing the stress-tensor. We also discuss why it is very likely that our results are not compatible with the imaginary central charge previously found for higher-spin theories in dS(3).}
\preprint{Nikhef-2012-019\\ICTS/2012/08}

\begin{document}

\baselineskip 3.5ex

\section{Introduction}
Since its proposal, the AdS/CFT duality \cite{Maldacena:1997re} has provided a very powerful paradigm for exploring quantum gravity in anti-de Sitter spacetimes. Recently, a particularly novel class of these dualities were found, which are the so-called higher-spin (HS)/CFT dualities. These are interesting for a variety of reasons, which we shall simply enumerate and refer the reader to the existing literature for details. Firstly, they provide non-supersymmetric examples of AdS/CFT with a very concrete proposal for a dual CFT \cite{Klebanov:2002ja, Gaberdiel:2010pz} without -- in many cases -- an explicitly known string embedding.\footnote{See \cite{Chang:2012kt} for an important exception.} Secondly, even for the more `stringy' dualities like \cite{Maldacena:1997re}, they have a particular relevance to the tensionless string (or equivalently, the free CFT) window of the dualitity. In particular, the twist-two sector of the free CFT is expected to be described by a higher-spin theory in AdS, and more generally, higher-spin symmetry may be expected to be present in the full tensionless string theory. See \cite{HaggiMani:2000ru} for details.

In the light of the first motivation, higher-spin theories in AdS$_3$ have been particularly interesting. Importantly, in three dimensions, higher-spin theories admit consistent truncations to theories containing a finite number of `higher' spins, an important simplification \textit{vis-a-vis} the usual situation in higher-dimensions. See \cite{Bekaert:2005vh} for a review of Vasiliev theories in general dimensions. A study of the asymptotic symmetry algebra \cite{BH} for these theories was carried out in \cite{Henneaux:2010xg,Campoleoni:2010zq} and this algebra was shown to be a $\WW$ algebra. This was further tested at the quantum level in \cite{Gaberdiel:2010ar} using the heat kernel methods of \cite{David:2009xg}.\footnote{See \cite{Gopakumar:2011qs} and \cite{Gupta:2012np} for an extension to higher-dimensional AdS spaces.} Importantly, from these inputs, a duality between these theories and minimal model CFTs was proposed in \cite{Gaberdiel:2010pz}. Since then, this subject has seen intense exploration, see for example \cite{Castro:2010ce} --\cite{Gaberdiel:2012uj} and we refer the reader to the recent review \cite{Gaberdiel:2012uj} and references therein for a very comprehensive account of these developments. Motivated by the conjecture \cite{Gaberdiel:2010pz}, an exploration of the holography of topologically massive higher-spin theories in AdS$_3$ was also initiated in \cite{Bagchi:2011vr} -- \cite{Chen:2012au}. 

Since the universe has a small positive cosmological constant, it is clearly very relevant to try and elucidate quantum gravity on de Sitter space. In particular, it would be very striking if we had concrete examples of de Sitter (dS)/CFT correspondences \cite{Strominger:2001pn}.\footnote{See \cite{Park:1998qk}--\cite{Park:1998yw} for early explorations in this regard.} In this light, dualities for higher-spin theories are an important avenue of exploration because firstly, higher-spin theories may just as well be defined on dS spaces as they are on AdS space\footnote{The essential ingredient that goes into the construction of Vasiliev higher spin theories is a non-zero cosmological constant \cite{Fradkin:1987ks,Vasiliev:1999ba}. Indeed, higher spin theories admitting de Sitter space as a vacuum solution were constructed in \cite{Iazeolla:2007wt,Iazeolla:2008bp}.} and there are already very concrete dualities available for these theories in AdS as mentioned above. Moreover, these dualities are explicitly non-supersymmetric and do not require an embedding into string theory. While the question of the string embedding and supersymmetry might still need to be addressed while formulating the eventual theory of our universe, it is important that these new avenues can potentially provide us with useful and computable models of quantum gravity on de Sitter space without having to directly deal with these formidable issues. Notably, an explicit realisation of the dS/CFT correspondence was proposed in \cite{Anninos:2011ui} where a CFT dual to a higher-spin theory in dS$_4$ was proposed by analytic continuation from the AdS case, which relates Vasiliev's higher spin theory in de Sitter space to a Euclidean $Sp(N)$ CFT$_3$. Subsequently there were some works which made the dictionary more precise and discussed subtleties in the programe opening up a plethora of open questions to be addressed in the future \cite{Ng:2012xp,Anninos:2012ft,Das:2012dt}.  In a subsequent development \cite{Ouyang:2011fs}, the analysis of \cite{Henneaux:2010xg,Campoleoni:2010zq} was extended to dS$_3$ space again by analytic continuation from AdS$_3$. In particular, the asymptotic symmetry algebra was computed for higher-spin theories in dS$_3$ alongwith the value of the Brown-Henneaux central charge. The central charge thus evaluated appeared to be an imaginary quantity.\footnote{See \cite{Park:1998qk},\cite{Balasubramanian:2001nb},\cite{Park:2007yq} for related work where an imaginary central charge appeared within the context of dS/CFT for pure gravity.}

In this paper, we shall use holographic renormalisation \cite{de Haro:2000xn} to compute correlation functions (in particular, two-point functions) in the dual CFT by computing AdS quantities. We are motivated by two reasons. Firstly, we shall explicitly show that the correlation functions in the dual CFT factorise holomorphically for both the stress energy tensor and the $\WW$ current. Secondly, we shall compute two point correlation functions in the dS/CFT case by analytic continuation of the AdS/CFT results. We find that the correlation functions thus arrived at are more likely to be compatible with the real central charge associated with dS$_3$ \cite{Strominger:2001pn} found by analyzing the stress tensor on $\hat{\II}^{-}$, which is the past infinity of the region accessible by a timelike observer, as opposed to the imaginary central charge found in \cite{Ouyang:2011fs}. We can say anything concrete only after the computation of the three point functions $\left<TWW\right>$, which we leave to address in the future. We mention here that there are a number of important conceptual and technical subtleties pertaining to dS/CFT, see for example \cite{Harlow:2011ke} which we shall not address here. We refer the reader to the references in \cite{Anninos:2011ui} for an account of progress in these questions.

A specific subtlety that is important for us pertains to the overall program of defining dS/CFT correlators by analytic continuation from AdS/CFT ones. When computing dS/CFT correlators in spacetimes that are only asymptotically de Sitter (or for operators other than the CFT stress tensor), a direct analytic continuation in the cosmological constant and Wick-rotating the AdS radial coordinate typically leads to inconsistent results. A more sophisticated continuation proposed in \cite{McFadden:2009fg,McFadden:2010na} leads to consistent results\footnote{We thank Marika Taylor for a very stimulating correspondence in this regard and also for bringing this work to our attention.}. A question that would be relevant to our analysis is if the direct analytic continuation that we carry out would lead to consistent correlation functions, even within the conformal vacuum. That we find this to be true, even at the preliminary level of a two-point function of a spin-3 theory is encouraging, but a stronger statement must surely await a fuller exploration of this question.\footnote{In general, as we have indicated previously, imaginary quantities do appear in dS/CFT, notably as conformal weights \cite{Balasubramanian:2001nb} or temperatures \cite{Bousso:2001mw} in the Kerr-de Sitter black hole. Our discussion is of course limited to correlators in the conformal vacuum. We thank Mu-In Park for discussions and correspondence regarding these points.}

Finally, the methods we employ here may reasonably be expected to extended to the topologically massive case in AdS$_3$ mentioned above. We expect to be able to compute two point functions in those theories to find concrete evidence of logarithmic behaviour as for topologically massive gravity in \cite{Skenderis:2009nt}. This is work in progress and we defer further discussions for the moment.

A brief overview of this paper is as follows. In Section \ref{bndy} we will begin with a brief review of the construction of the boundary action from \cite{Joung:2011xb}. We will make the section self contained as far as the notations of \cite{Joung:2011xb} are concerned. We would like to stress that the most crucial aspects of our computation would be the anomaly equations (\refb{bdy17} for AdS and \refb{ds8} for dS), since pure higher spin theory in 2+1 dimensions has no propagating degrees of freedom, and hence would be pure gauge. Hence the computation of on-shell action is a simple computation of the anomaly. In Section \ref{holren} we will fix the counter term action by holographic renormalization technique and compute the two point correlators. Two point functions are very much normalization dependent. We can absorb any overall factors and that can change the overall coefficient. However, once we fix our choice of normalization, we should get three point functions which are consistent with the asymptotic symmetry algebra \cite{Campoleoni:2010zq}. We will discuss this consistency requirement on the three point correlators $\left<TWW\right>$ at the end of the section which we leave for a future work. In Section \ref{ds} we will analytically continue our results for Euclidean AdS$_3$ from the previous sections to compute the $\left<WW\right>$ and $\left<TT\right>$correlators for a spin-3 theory coupled to de-Sitter gravity on the past infinity $\hat{\II}^{-}$ of a region $\OO^{-}$ which comprises the causal past of a timelike observer. We find that our results are more likely to be compatible with the real central charge associated with dS$_3$ \cite{Strominger:2001pn} found by analyzing the stress tensor on $\hat{\II}^{-}$ as opposed to the imaginary central charge found in \cite{Ouyang:2011fs}. There is however a consistency requirement, like the AdS case, on the three point functions, which we hope to address in the future. We conclude in the next section.

\section{Boundary Action for the Spin-3 Theory in AdS$_3$}\label{bndy}
In this section, we will review the boundary action of free AdS higher spin fields derived in \cite{Joung:2011xb}. We  restrict ourselves to spin-2 and spin-3 fields\footnote{By which we mean completely symmetric rank-3 and rank-2 tensors, the spin-2 field is, of course, the graviton.} and $2+1$ dimensional Euclidean AdS$_3$ space. For generic spins and dimensions, we refer the reader to \cite{Joung:2011xb}. This section is also a self-contained review of the notations of \cite{Joung:2011xb}, though we have reinserted the AdS radius $\ell$ and the Newton's constant at appropriate places. 

The Euclidean AdS$_3$ metric in the Poincar\'e patch is given by
\be{bdy}
ds^2={\ell^2\over\sigma^2}\left(d\sigma^2+\delta_{ij}dx^{i}dx^{j}\right),
\ee
where $\delta=diag(1,1)$. We can also write this metric in terms of holomorphic coordinates $(z,\bar{z})$ as
\be{bdy1}
ds^2={\ell^2\over\sigma^2}\left(d\sigma^2+dz d\bar{z}\right),
\ee
where $z=x_1+ix_2$. We define our integration measure and delta function (to be used later), as below
\be{bndy1}
d^2 z \equiv dx^1 dx^2, \quad \delta^{(2)}\left(\vec{z}\right)\equiv \delta\left(x^1\right)\delta\left(x^2\right)
\ee
To avoid proliferation of indices the higher spin fields are contracted with auxiliary variables $U^{A}\in R^{3}$ and vielbeins $\bar{e}^{M}_{A}={\sigma\over \ell}\delta^{M}_{A}$, as
\ben{bdy2}
\Phi^{(3)}(x,\sigma,U) &\equiv& {1\over 6}\left(U^{A}U^{B}U^{C}\right)\left(\bar{e}^{M}_{A}\bar{e}^{N}_{B}\bar{e}^{P}_{C}\right)\Phi_{MNP,} \nonumber \\
\Phi^{(2)}(x,\sigma,U) &\equiv& {1\over 2}\left(U^{A}U^{B}\right)\left(\bar{e}^{M}_{A}\bar{e}^{N}_{B}\right)\Phi_{MN}.
\een
Note that since dimension of $\Phi_{MNP}$ is $\ell^3$ and $\Phi_{MN}$ is $\ell^{2}$, $\Phi^{(3)}$ and $\Phi^{(2)}$ becomes dimensionless. We also define the covariant derivative and Fronsdal operator in terms of the auxiliary variables $U$ as\footnote{We have slight differences with \cite{Joung:2011xb} because of difference in some conventions. We will make our conventions clear.}
\ben{bdy3}
\DD_{A} &=&\bar{e}_{A}^{M}\p_{M}-\bar{e}_{A}^{M}\omega_{MB}^{~~~C}U^{B}\p_{U^C}={\sigma\over \ell}\p_{A}-{1\over \ell}\left[U_A \p_{U^{\sigma}}-U_{\sigma}\p_{U^{A}}\right] ,\nonumber \\
\FF^{(s)} &=&\left(\p_{U}.\DD\right)\left(U.\DD\right)-2\left(U.\DD\right)\left(\p_{U}.\DD\right)+{1\over 2}\left(U.\DD\right)^{2}\p_{U}^{2}-{(s+1)\over 2\ell^2}U^2\p_{U}^{2}+{4s\over \ell^2}.
\een
We have defined the Fronsdal operator $\FF^{(s)}$ for generic spin-s in 2+1 dimensions, but we will only be interested in $s=2$ and $3$. The spin connection $\omega_{M}^{~AB}$ is defined as\footnote{We define our symmetric brackets `(\ldots)' and anti-symmetric brackets `[\ldots]' as the minimum (signed) permutations in the enclosed indices required without any normalization factor.}
\be{bdy4}
\omega_{M}^{AB}={1\over 2}e^{N[A}\p_{M}e_{N}^{B]}-{1\over 2}\Gamma_{MN}^{~~P}e_{P}^{[A}e^{B]N},
\ee
where $\Gamma_{MN}^{~~P}$ is the Christoffel connection defined as
\be{bdy5}
\Gamma_{MN}^{~~P}={1\over 2}g^{PQ}\left(\p_{(M}g_{N)Q}-\p_{Q}g_{MN}\right)
\ee
One can easily check from the various definitions above that
\ben{bdy6}
(\FF\Phi)^{(3)}_{MNP} &\equiv& \FF_{MNP} =\nabla^{2}\Phi_{MNP}-\nabla_{(M}\nabla^{Q}\Phi_{NP)Q}+{1\over 2}\nabla_{(M}\nabla_{N}\Phi_{P)Q}^{~~~Q}-{2\over \ell^2}\bar{g}_{(MN}\phi_{P)Q}^{~~~Q} \nonumber \\
(\FF\Phi)^{(2)}_{MN} &\equiv& \FF_{MN} =\nabla^{2}\Phi_{MN}-\nabla_{(M}\nabla^{Q}\Phi_{N)Q}+{1\over 2}\nabla_{(M}\nabla_{N)}\Phi_{Q}^{~Q}-{2\over \ell^2}\bar{g}_{MN}\phi_{Q}^{~Q}+{2\over \ell^{2}}\phi_{MN} \nonumber \\
\een
The action for the spin-3 field coupled to background $AdS$ gravity is given by \footnote{This is the usual Fronsdal action for spin 3 field \cite{Fronsdal:1978rb}. See also \cite{Buchbinder:2001bs,Sagnotti:2003qa}.}
\ben{bdy7}
S^{(3)} &=& {1\over 64\pi G_{N}}\int_{\sigma=\sigma_B}^{\sigma=\infty} d^3 x ~ \sqrt{\bar{g}}\Phi^{MNP}\left(\FF_{MNP}-{1\over 2}\bar{g}_{(MN}\FF_{P)}\right) \nonumber \\
&=& {3\ell^3 \over 32\pi G_N}\int_{\sigma_B}^{\infty}{d\sigma\over \sigma^{3}}\langle\langle\Phi^{(3)}\left|\left(1-{1\over4}U^2\p_{U}^{2}\right)\FF^{(3)}\Phi^{(3)}\right.\rangle\rangle.\een 
The action for the spin-2 field is 
\ben{bdy7b}
S^{(2)} &=& {1\over 64 \pi G_{N}}\int_{\sigma=\sigma_B}^{\sigma=\infty} d^3 x ~ \sqrt{\bar{g}}\Phi^{MN}\left(\FF_{MN}^{(2)}-{1\over 2}\bar{g}_{MN}\FF^{(2)}\right) \nonumber \\
&=& {\ell^3 \over 32\pi G_N}\int_{\sigma_B}^{\infty}{d\sigma\over \sigma^{3}}\langle\langle\Phi^{(2)}\left|\left(1-{1\over4}U^2\p_{U}^{2}\right)\FF^{(2)}\Phi^{(2)}\right.\rangle\rangle.
\een
These differ from \cite{Joung:2011xb} by a sign and upto a choice of normalization $1\over 64\pi G_N$. The double angular brackets $\left<\left<.\left|.\right.\right>\right>$ are defined as
\be{bdy8}
\left<\left<f\left|g\right.\right>\right>\equiv \int d^2 z ~\sum_{n=0}^{\infty}{1\over n!}f_{m_1\cdots m_n}g^{m_1 \cdots m_n}.
\ee
The Fronsdal equations $\FF\Phi=0$, have the gauge invariance
\ben{bdy9}
\delta \Phi^{(3)} (x,\sigma,U)&=& U.\DD\varepsilon^{(3)}(x,\sigma,U), \nonumber \\
\delta \Phi^{(2)}(x,\sigma,U) &=& U.\DD\varepsilon^{(2)}(x,\sigma,U).
\een
where the gauge parameter $\varepsilon^{(3)} (x,\sigma,U)$ is subject to the traceless constraint $\p_{U}^{2}\varepsilon^{(3)}=0$. The action is also invariant under the above gauge transformation upto terms which do not vanish at the boundary. The boundary action was derived \cite{Joung:2011xb}, by demanding the restoration of the invariance under gauge transformations which do not vanish at the boundary. It was found that not all gauge invariance could be restored and that a subset of the gauge transformation still remains anomalous. We will later see that this anomaly will be crucial in giving a non zero on-shell action and hence finite non-zero correlation functions.

In what follows, it is convenient to decompose the spin-3 and spin-2  field $\Phi$ and gauge parameter $\varepsilon$, in terms of their boundary spins. Let us write $U^{\sigma}=v$ and $U^{i}=u^{i}$. The decomposition is
\ben{bdy10}
\Phi^{(3)}(x,\sigma;u,v)\equiv \sum_{r=0}^{3}{v^r \over r!}\phi^{(3-r)}_{3}(x,\sigma;u)&,&
\varepsilon^{(3)}(x,\sigma;u,v)\equiv \sum_{r=0}^{2}{v^r \over r!}\epsilon^{(2-r)}_{3}(x,\sigma;u), \nonumber \\
\Phi^{(2)}(x,\sigma;u,v)\equiv \sum_{r=0}^{2}{v^r \over r!}\phi^{(2-r)}_{2}(x,\sigma;u)&,&
\varepsilon^{(2)}(x,\sigma;u,v)\equiv \sum_{r=0}^{1}{v^r \over r!}\epsilon^{(1-r)}_{2}(x,\sigma;u). \nonumber \\
\een
The subscripts on the R.H.S keep track of which field decomposition they refer to and the superscripts are the boundary spin. In the above sum, tracelessness of $\varepsilon^{(3)}$ relates $\epsilon^{(0)}_{3}=-\p_{u}^{2}\epsilon^{(2)}_{3}$. In terms of the above decomposition the gauge transformation of spin-3 field \refb{bdy9}, becomes 
\ben{bdy11}
\delta\phi^{(3)}_{3}&=& {1\over \ell}\left[\sigma\left(u.\p\right)\epsilon^{(2)}_{3}-u^{2}\epsilon^{(1)}_{3}\right], \nonumber \\
\delta\phi^{(2)}_{3}&=& {1\over \ell}\left[\left(\sigma\p_{\sigma}+2+u^2\p_{u}^{2}\right)\epsilon^{(2)}_{3}+\sigma\left(u.\p\right)\epsilon^{(1)}_{3}\right], \nonumber \\
\delta\phi^{(1)}_{3}&=& {1\over \ell}\left[-\sigma\left(u.\p\right)\p_{u}^{2}\epsilon^{(2)}_{3}+2\left(\sigma\p_{\sigma}+1\right)\epsilon^{(1)}_{3}\right], \nonumber \\
\delta\phi^{(0)}_{3}&=& -{3\over \ell}\sigma\p_{\sigma}\p_{u}^{2}\epsilon^{(2)}_{3}.
\een
The corresponding gauge transformations for spin-2 read
\ben{bdy111}
\delta\phi^{(2)}_{2}&=& {1\over \ell}\left[\sigma\left(u.\p\right)\epsilon^{(1)}_{2}-u^{2}\epsilon^{(0)}_{2}\right], \nonumber \\
\delta\phi^{(1)}_{2}&=& {1\over \ell}\left[\left(\sigma\p_{\sigma}+1\right)\epsilon^{(1)}_{2}+\sigma\left(u.\p\right)\epsilon^{(0)}_{2}\right], \nonumber \\
\delta\phi^{(0)}_{2}&=& {2\over \ell}\sigma\p_{\sigma}\epsilon^{(0)}_{2}.
\een
The construction of the boundary action in terms of the boundary decomposition proceeds as follows. First, the requirement of a well defined variational principle under Dirichlet boundary conditions gives terms depending on radial derivatives of the field at the boundary. Then several terms not depending on the radial derivative are added to the action demanding that the gauge invariance with non-vanishing gauge parameters at the boundary is restored. It was found that gauge invariance under non vanishing $\epsilon^{(2)}_{3}$, $\p_{\sigma}\epsilon^{(2)}_{3}$, $\p_{\sigma}\epsilon^{(1)}_{3}$, $\epsilon^{(1)}_{2}$, $\p_{\sigma}\epsilon^{(1)}_{2}$ and $\p_{\sigma}\epsilon^{(0)}_{2}$ could be restored, but a non-vanishing $\epsilon^{(1)}_{3}$ and $\epsilon^{(0)}_{2}$ remains anomalous. The boundary action obtained in \cite{Joung:2011xb}, by following these steps is
\ben{bdy12}
S_{bdy}^{(3)}&=& -{3\sigma_{B}^{-2}\ell\over 32\pi G_N}\left[\left<\left<\phi^{(3)}_{3}\left|\left(-1+{1\over 2}u^2\p_{u}^{2}\right)\chi^{(3)}_{3}\right.\right>\right>_{\sigma_{B}}+{1\over 6}\left<\left<3\p_{u}^{2}\phi^{(2)}_{3}+\phi^{(0)}_{3}\left|\chi^{(0)}_{3}\right.\right>\right>_{\sigma_{B}} \right. \nonumber \\
&& +\left. {1\over 2}\left<\left<\p_{u}^{2}\phi^{(3)}_{3}-\phi^{(1)}_{3}\left|\zeta^{(1)}_{3}\right.\right>\right>_{\sigma_{B}}+{1\over 18}\left<\left<\zeta^{(1)}_{3}\left|\zeta^{(1)}_{3}\right.\right>\right>_{\sigma_{B}}\right], \nonumber \\
S_{bdy}^{(2)} &=& -{\ell\sigma_{B}^{-2}\over 32\pi G_N}\langle\langle\phi_{2}^{(2)}\left|\left(-1+{1\over 2}u^2 \p_{u}^{2}\right)\chi_{2}^{(2)}\right.\rangle\rangle_{\sigma_{B}},
\een
where we have used
\ben{bdy13}
\chi^{(3)}_{3}&\equiv& \p_{\sigma}\left(\sigma\phi^{(3)}_{3}\right)-\sigma\left(u.\p\right)\phi^{(2)}_{3}-u^2\phi^{(1)}_{3}, \nonumber \\
\chi^{(0)}_{3}&\equiv& {1\over 2}\sigma\p_{\sigma}\left(3\p_{u}^{2}\phi^{(2)}_{3}+\phi^{(0)}_{3}\right)+3\phi^{(0)}_{3}, \nonumber \\
\zeta^{(1)}_{3}&\equiv& {1\over 2}\sigma u.\p\left(3\p_{u}^{2}\phi^{(2)}_{3}+\phi^{(0)}_{3}\right)+9\phi^{(1)}_{3}. \nonumber \\
\chi_{2}^{(2)}&\equiv& \sigma\p_{\sigma}\phi_{2}^{(2)}-\sigma\left(u.\p\right)\phi_{2}^{(1)}+{1\over 2}u^2\phi_{2}^{(0)}.
\een
One can add another term to the boundary action of the spin-3 field which is invariant under $\epsilon^{(2)}_{3}$, $\p_{\sigma}\epsilon^{(2)}_{3}$ and $\p_{\sigma}\epsilon^{(1)}_{3}$ but anomalous under $\epsilon_{3}^{(1)}$, which is
\be{bdy14}
S_{bdy}^{(CT)}= -{3\sigma_{B}^{-2}\ell\over 16\pi G_N}\left[\left<\left<\phi^{(3)}_{3}\left|G^{(3)}\right.\right>\right>_{\sigma_{B}}+{1\over 18}\left<\left<\left(\phi^{(0)}_{3}+3\p_{u}^{2}\phi^{(2)}_{3}\right)\left|K^{(0)}\right.\right>\right>_{\sigma_{B}}\right],
\ee
where
\ben{bdy15}
G^{(3)}&=& c\left(1-{1\over 4}u^2\p_{u}^{2}\right)A^{(3)}, \nonumber \\
K^{(0)}&=& -c{\sigma\over 4}\left(\p.\p_{u}\right)\p_{u}^{2}A^{(3)}, \nonumber \\
A^{(3)} &=& \sigma^{2}\FF_{2}\phi^{(3)}-{1\over 36}\sigma^{3}\left(u.\p\right)^{3}\left(\phi^{(0)}+3\p_{u}^{2}\phi^{2}\right),
\een
where $c$ is an arbitrary coefficient and $\FF_{2}$ is the 2-d flat-Fronsdal operator
\be{bdy16}
\FF_{2}\equiv \p^{2}-u.\p \p_{u}.\p +{1\over 2}\left(u.\p\right)^{2}\p_{u}^{2}.
\ee
One can also add a similar term to the boundary action of the spin-2 field but it turns out that this term is anomaly free for $d=2$ (see equation (2.42) and (2.43) of \cite{Joung:2011xb}) and hence will vanish on-shell, since our solutions will be pure gauge. As mentioned before, this action $S+S_{bdy}+S_{bdy}^{(CT)}$ is anomalous under a non-vanishing $\epsilon^{(1)}_{3}$ and $\epsilon_{2}^{(0)}$ at the boundary, and the anomaly is given by \cite{Joung:2011xb}
\ben{bdy17}
\delta_{\epsilon^{(1)}_{3}}\left(S^{(3)}+S_{bdy}^{(3)}+S_{bdy}^{(CT)}\right) &=&-{3\over 32\pi G_N}\left<\left<\epsilon^{(1)}\left|\AAA^{(1)}_{3}\right. \right>\right>, \nonumber \\
\delta_{\epsilon^{(0)}_{2}}\left(S^{(2)}+S_{bdy}^{(2)}\right) &=& {1\over 32\pi G_N}\langle\langle\epsilon_{2}^{(0)}\left|\AAA^{(0)}_{2}\right.\rangle\rangle,
\een
where we have defined
\ben{bdy18}
\AAA^{(1)}_{3}&\equiv& \sigma^{-2}\left[-{1\over 2}+2c-c{\sigma^{2}\over 6}u.\p \p_{u}.\p\right] \p_{u}^{2}A^{(3)}, \nonumber \\
\AAA^{(0)}_{2} &\equiv& \left[\p^{2}\p_{u}^{2}-\left(\p_{u}.\p\right)^{2}\right]\phi_{2}^{(2)}.
\een
In the next section, we will use holographic renormalization to fix the arbitrary coefficient $c$ and then compute the two point correlators from the finite part of the on-shell action via AdS/CFT correspondence.
\section{Holographic Renormalization  in AdS$_3$ and Two-point Correlators}\label{holren}
In the semi-classical (planar) limit of the AdS/CFT correspondence, $e^{-S^{(tot)}_{on-shell}}$ is the generating functional for correlators of boundary currents which couple to the leading behavior of the bulk fields \cite{Witten:1998qj,Gubser:1998bc}, where $S^{(tot)}=S+S_{bdy}+S_{bdy}^{(CT)}$. Typically the on-shell action is divergent. To regulate the divergence, the boundary is kept at a non-zero cut-off $\sigma_{B}$ and the terms which are divergent as $\sigma_{B}\to 0$ are cancelled by appropriately chosing the arbitrary constant $c$ in the counter-term action. Finally, the cut-off $\sigma_{B}$ is sent to zero. This is the principle of holographic renormalization \cite{de Haro:2000xn}. 

The first step in this procedure is to obtain the solution to the equations of motion $\FF\Phi=0$. The solutions for generic spins $s$ and dimensions $d+1$ , are obtained in \cite{Joung:2011xb} as a gauge transformation over a truncated solution
\be{hr1}
\phi^{(s-1)}=\sigma^{2-s}U_{d+2s-4\over 2}\left(q\sigma\right)h_{TT}^{(s)}(x), \quad \phi^{(s-1)}=\phi^{(s-2)}=\phi^{(s-3)}=0.
\ee
where $q=\sqrt{-\p^{2}}$ and $h_{(TT)}^{(s)}(x)$ is subject to the transverse, traceless constraint
\be{hr2}
\p_{u}^{2}h_{TT}^{(s)}=0, \quad \p_{u}.\p h_{TT}^{(s)}=0,
\ee
and $U_{n}(z)$ is related to the modified Bessel function $K_{n}(z)$ as
\be{hr3}
U_{n}(z)\equiv {2\over \Gamma(n)}\left({z\over 2}\right)^{n}K_{n}(z).
\ee
The transverse traceless projection is given as 
\be{hr4}
h_{TT}^{(s)}=h^{(s)}-u.\p\bar{\rho}^{(s-1)}[h^s]+u^2\rho^{(s-2)}[h^s],
\ee
where $\bar{\rho}^{(s-1)}[h^s]$ and $\rho^{(s-2)}[h^s]$ are non-local functionals of $h^{(s)}$, and $\bar{\rho}^{(s-1)}[h^s]$ is traceless i.e $\p_{u}^{2}\bar{\rho}^{(s-1)}[h^s]=0$. For $s=3$ and $s=2$, we present the component form of $\bar{\rho}^{(2)}_{3}[h^3]$, $\rho^{(1)}_{3}[h^3]$, $\bar{\rho}_{2}^{(1)}[h^2]$ and $\rho_{2}^{(0)}[h^2]$ which is
\ben{hr5a}
\left(\bar{\rho}^{(2)}_{3}\right)_{jk} = \alpha_{jk}-{1\over 2}\delta_{jk}\delta^{ij}\alpha_{ij}, \quad
\left(\rho^{(1)}_{3}\right)_{k} = -{1\over 6}h_{k}-{1\over 12}{\p_{k}\p^{l}h_{l}\over \p^{2}}+{\p^{l}\p^{m}h_{lmk}\over 6 \p^{2}},\een
where \ben{hr5b}
\alpha_{ij} = {\p^{l}h_{ijl}\over \p^2}-{\p_{(i}\p^{l}\p^{m}h_{j)lm}\over 3\p^4}-{\p_{(i}h_{j)}\over 6\p^{2}}+{\p_{i}\p_{j}\p^{l}h_{l}\over 3\p^4},\een
and finally, \ben{hr5c}
\left(\bar{\rho}_{2}^{(1)}\right)_{k}= {\p^{l}h_{kl}\over \p^{2}}-{1\over 2}{\p_{k}h\over \p^{2}}, \quad
\left(\rho^{(0)}_{2}\right)= {\p^{i}\p^{j}h_{ij}\over 2\p^2}-{1\over 2}h.
\een
The full solution is a gauge transformation on the above truncated solution with gauge transformation parameter
\ben{hr6}
\epsilon^{(s-1)}&=& \ell\sigma^{1-s}U_{d+2s-4\over 2}(q\sigma)\bar{\rho}^{(s-1)}[h^{(s)}], \nonumber \\
\epsilon^{(s-2)} &=& \ell\sigma^{2-s}U_{d+2s-4\over 2}(q\sigma)\rho^{(s-2)}[h^(s)].
\een
Let us now note that $h_{TT}^{(s)}=0$ for $d=2$ for any spin-$s$\footnote{This can be easily seen by counting the number of components of $h^{(s)}$ and the number of components the transverse traceless projection projects out. They turn out to be equal for $d=2$ and hence $h_{TT}^{(s)}=0$ for $d=2$.}. This means that for our case, the solution is a pure gauge with the gauge transformation parameters \refb{hr6} in which $s$ is restricted to 3 and 2 and $d$ is restricted to 2. Let us write down the complete solution for $s=3,2$ and $d=2$.\footnote{See \cite{Joung:2011xb} for the generic solution.}
\ben{hr7}
\phi^{(3)}_{3}= \sigma^{-1}U_2 \left(q\sigma \right)h^{(3)}_{3}, &&\left| ~\phi_{2}^{(2)}=U_{1}\left(q\sigma\right)h^{(2)}_{2} \right. ,\nonumber \\
\phi^{(2)}_{3} = -{q^2\over 2}U_{1}\left(q\sigma\right)\bar{\rho}^{(2)}+U_2\left(q\sigma\right)u.\p \rho^{(1)}, &&\left| ~ \phi^{(1)}_{2}=-q^2\sigma K_{0}\left(q\sigma\right)\bar{\rho}^{(1)}_{2}+\sigma U_{1}\left(q\sigma\right)\left(u.\p\right)\rho^{(0)}_{2} \right. ,\nonumber \\
\phi^{(1)}_{3} = -q^2\sigma U_1\left(q\sigma\right)\rho^{(1)}, \quad \phi^{(0)}=0, && \left| ~ \phi^{(0)}_{2}=-2q^2\sigma^{2}K_{0}\left(q\sigma\right)\rho^{(0)}_{2} \right. ,
\een
Since our solution is a pure gauge, all that we need to know for the on-shell action is $\epsilon^{(1)}_{3}$, $\epsilon_{2}^{(0)}$, $\AAA^{(0)}_{2}$ and $\AAA^{(1)}_{3}$, which are (from \ref{bdy18}, \ref{bdy15}, \ref{hr6}, \ref{hr7})
\ben{hr8}
\epsilon^{(1)}_{3}&=&  \ell\sigma_{B}^{-1}U_{2}(q\sigma_{B})\rho^{(1)}[h^{(3)}], \nonumber \\
\epsilon^{(0)}_{2}&=& \ell U_{1}\left(q\sigma\right)\rho^{(0)}_{2}, \nonumber \\
\AAA^{(1)}_{3} &=& \sigma_{B}^{-1}\left[-{1\over 2}+2c\right]U_{2}\left(q\sigma_{B}\right)\p_{u}^{2}\FF_{2} h^{(3)} -\sigma_{B}\left[-{1\over 2}+2c\right]U_{2}\left(q\sigma_{B}\right)\p^{2}\left(u.\p\right)\left(\p_{u}.\p\right)\rho^{(1)}, \nonumber \\
&& -{c\over 6}\sigma_{B} U_{2}\left(q\sigma_{B}\right)\left(u.\p\right)\left(\p_{u}.\p\right)\p_{u}^{2}\FF_{2} h^{(3)}+{c\over 6}\sigma_{B}^{3}U_{2}\left(q\sigma_{B}\right)\p^{4}\left(u.\p\right)\left(\p_{u}.\p\right)\rho^{(1)}, \nonumber \\
\AAA^{(0)}_{2}&=& U_{1}\left(q\sigma\right)\left[\p^{2}\p_{u}^{2}-\left(\p_{u}.\p\right)^{2}\right]h^{(2)}.
\een
It is clear from above that the on-shell action \refb{bdy17} will have an $\OO(\sigma_{B}^{-2})$ divergence unless the first term in $\AAA^{(1)}_{3}$ does not go to zero. This fixes $c$ to be ${1\over4}$, and the on-shell action becomes
\be{hr9}
S_{on-shell}^{tot}= S^{(3)}+S^{(2)},
\ee
where,
\ben{hr91}
S^{(3)}&=& -{3\over 32\pi G_N}\left<\left<\epsilon^{(1)}_{3}\left|\AAA^{(1)}_{3}\right. \right>\right> \nonumber \\
&=& -{\ell\over 256\pi G_N}\int d^2 z~ \p^{k}\left(\rho^{(1)}_{3}\right)_{k}U_2\left(q\sigma\right)U_2\left(q\sigma\right)\left[3\p^{2}\p^{l}h_{l}-2\p^{k}\p^{l}\p^{m}h_{lmk}\right]+\OO(\sigma_{B}^{2}) \nonumber \\
&=& -{\ell\over 256\pi G_N}\int d^2 z~ \p^{k}\left(\rho^{(1)}_{3}\right)_{k}\left[3\p^{2}\p^{l}h_{l}-2\p^{k}\p^{l}\p^{m}h_{lmk}\right]+\OO(\sigma_{B}^{2}) \nonumber \\
&=& -{\ell\over 768\pi G_N}\int d^2 z~ h^{ijk}\left({\p_{i}\p_{j}\p_{k}\p_{l}\p_{m}\p_{n}\over \p^{2}}\right)h^{lmn}+local+\OO(\sigma_{B}^{2}) \nonumber \\
&=& -{\ell\over 48\pi G_N}\int d^2 z\left[h_{zzz}\left({\bar{\p}^{5}\over \p}\right)h_{zzz}+h_{\bar{z}\bar{z}\bar{z}}\left({{\p}^{5}\over \bar{\p}}\right)h_{\bar{z}\bar{z}\bar{z}}\right] +local+\OO(\sigma_{B}^{2}),
\een
and
\ben{hr92}
S^{(2)}&=& {1\over 32\pi G_N}\langle\langle\epsilon_{2}^{(0)}\left|\AAA_{2}^{(0)}\right.\rangle\rangle \nonumber \\
&=& {\ell\over 32\pi G_{N}}\int d^2 z~ \rho_{2}^{(0)}U_{1}(q\sigma)U_{1}(q\sigma)\left[\p^{2}h-\p^{l}\p^{m}h_{lm}\right] \nonumber \\
&=& -{\ell\over 64\pi G_N}\int d^2 z ~ h^{ij}\left(\p_{i}\p_{j}\p_{l}\p_{m}\over \p^{2}\right)h^{lm} +local+\OO(\sigma_{B}^{2}) \nonumber \\
&=& {-\ell\over 16\pi G_{N}}\int d^2 z~ \left[h_{zz}\left(\bar{\p}^{3}\over \p\right)h_{zz}+h_{\bar{z}\bar{z}}\left({\p}^{3}\over \bar{\p}\right)h_{\bar{z}\bar{z}}\right]+local+\OO(\sigma_{B}^{2}).
\een
In the second line of the above two equations, we have used the asymptotic expansion of $U_{n}(z)$, which is 
\be{hr10}
U_{n}(z)=1+\OO(z^2)
\ee
In the last lines of the above two equations $\p\equiv {\p\over \p z}$ and $\bar{\p}\equiv {\p\over \p\bar{z}}$. We can use the identity \cite{Skenderis:2009nt},
\be{hr11}
{1\over \p\bar{\p}}\delta^{2}\left(\vec{z}-\vec{w}\right)={1\over 4\pi}\log\left(m^2\left|z-w\right|^{2}\right).
\ee
to write the on-shell action \refb{hr9} as (upto local and $\OO(\sigma_{B}^{2})$ terms)
\ben{hr12}
S_{on-shell}^{tot} &=& {5\ell\over 8\pi^{2} G_N}\int \int d^2 z~ d^2 w\left[{h_{zzz}(\vec{z})h_{zzz}(\vec{w})\over\left(\bar{z}-\bar{w}\right)^{6}}+{h_{\bar{z}\bar{z}\bar{z}}(\vec{z})h_{\bar{z}\bar{z}\bar{z}}(\vec{w})\over\left(z-w\right)^{6}}\right] \nonumber \\
&& +\, {3\ell\over 32\pi^{2} G_N}\int \int d^2 z~ d^2 w\left[{h_{zz}(\vec{z})h_{zz}(\vec{w})\over\left(\bar{z}-\bar{w}\right)^{4}}+{h_{\bar{z}\bar{z}}(\vec{z})h_{\bar{z}\bar{z}}(\vec{w})\over\left(z-w\right)^{4}}\right].
\een
The AdS-CFT conjecture implies that \cite{Witten:1998qj,Gubser:1998bc},
\be{hr13}
\left<\OO_{1}\left(\vec{z_1}\right)\cdots \OO_{n}\left(\vec{z_n}\right)\right>\equiv \left(8i\pi\right)^{n}\left(\delta^{n}\over\delta\Phi_1 \left(\vec{z_1}\right)\cdots \Phi_1 \left(\vec{z_1}\right)\right)e^{-S_{on-shell}}
\ee
Where, $\Phi_{1}$ is the bulk field dual to operator $\OO_{1}$ at the boundary. The normalization $(8\pi i)^{n}$, is a choice of normalization and is chosen such that $\left<T_{zz}(z)T_{zz}(w)\right>={c_{BH}\over 2\left(z-w\right)^{2}}$ (as we will verify below), where $c_{BH}={3\ell\over 2G_N}$ is the Brown-Hennaux central charge \cite{BH}\footnote{The choice of an $i$ in the normalization is to make the two-point function positive, and has also been used before in \cite{David:2007ak}, where the overall normalization used in defining the n-point function is $(i\pi)^{n}$. Once we fix the normalization at the level of two point function, the three point function should be able to tell us whether our central charge is positive, negative or imaginary as we will discuss later.}. Hence
\ben{hr14}
\left<W_{zzz}(\vec{z})W_{zzz}(\vec{w})\right>&=& (8\pi i)^2{\delta^{2}\over \delta h^{zzz}(\vec{z})\delta h^{zzz}(\vec{w})}e^{-S_{on-shell}^{tot}}{|_{h_{zzz}=h_{\bar{z}\bar{z}\bar{z}}=h_{zz}=h_{\bar{z}\bar{z}\bar{z}}=0}} \nonumber \\
&=& {5\ell\over 4G_N \left(z-w\right)^{6}}={5 c_{BH}\over 6 \left(z-w\right)^{6}}, \nonumber \\
\left<W_{\bar{z}\bar{z}\bar{z}}(\vec{z})W_{\bar{z}\bar{z}\bar{z}}(\vec{w})\right>&=& (8\pi i)^2{\delta^{2}\over \delta h^{\bar{z}\bar{z}\bar{z}}(\vec{z})\delta h^{\bar{z}\bar{z}\bar{z}}(\vec{w})}e^{-S_{on-shell}^{tot}}{|_{h_{zzz}=h_{\bar{z}\bar{z}\bar{z}}=h_{zz}=h_{\bar{z}\bar{z}}=0}} \nonumber \\
&=& {5\ell\over 4G_N \left(\bar{z}-\bar{w}\right)^{6}} = {5 c_{BH}\over 6 \left(\bar{z}-\bar{w}\right)^{6}}, \nonumber \\
\left<T_{\bar{z}\bar{z}}(\vec{z})T_{\bar{z}\bar{z}}(\vec{w})\right>&=& (8\pi i)^2{\delta^{2}\over \delta h^{\bar{z}\bar{z}}(\vec{z})\delta h^{\bar{z}\bar{z}}(\vec{w})}e^{-S_{on-shell}^{tot}}{|_{h_{zzz}=h_{\bar{z}\bar{z}\bar{z}}=h_{zz}=h_{\bar{z}\bar{z}}=0}} \nonumber \\
&=& {3\ell\over 4G_N \left(\bar{z}-\bar{w}\right)^{4}}= {c_{BH}\over 2 \left(\bar{z}-\bar{w}\right)^{4}} ,\nonumber \\
\left<T_{zz}(\vec{z})T_{zz}(\vec{w})\right>&=& (8\pi i)^2{\delta^{2}\over \delta h^{zz}(z)\delta h^{zz}(w)}e^{-S_{on-shell}^{tot}}{|_{h_{zzz}=h_{\bar{z}\bar{z}\bar{z}}=h_{zz}=h_{\bar{z}\bar{z}}=0}} \nonumber \\
&=& {3\ell\over 4G_N\left(z-w\right)^{4}}={c_{BH}\over 2 \left(z-w\right)^{4}}.
\een
The above results are consistent with the Euclidean CFT expectations, but the normalization is different from the $\WW$-algebra in \cite{Bouwknegt:1992wg}, which gives $c\over 3$ instead of $5c\over 6$. However, as we have argued in the introduction, two point functions are very much normalization dependent. Any overall factors can be absorbed by redefining the sources which is equivalent to redefining the currents in the boundary. Therefore, two point functions are not sufficient to show consistency with the asymoptotic symmetry algebra. We also need information from the three point functions. We fix the normalization such that the two point functions has positive real coefficient as in \refb{hr13}, and below we discuss the consistency requirement on the three point functions and associated central charge. 
\ben{hr15}
 \left<T(z)T(w)T(v)\right>&=&{\pm c_{BH}\over(z-w)^{2}(w-v)^{2}(v-z)^{2}}, \nonumber \\ \left<T(z)W(w)W(v)\right>&=& {\pm 5c_{BH}\over 2(z-w)^{2}(w-v)^{4}(v-z)^{2}}, \quad  
c = c_{BH}
\een
\ben{hr16}
 \left<T(z)T(w)T(v)\right> &=& {\pm ic_{BH}\over(z-w)^{2}(w-v)^{2}(v-z)^{2}}, \nonumber \\ \left<T(z)W(w)W(v)\right>&=&{\pm 5ic_{BH}\over 2(z-w)^{2}(w-v)^{4}(v-z)^{2}}, \quad
c = -c_{BH}
\een
\ben{hr17}
 \left<T(z)T(w)T(v)\right> &=&{\pm \sqrt{i} c_{BH}\over(z-w)^{2}(w-v)^{2}(v-z)^{2}}, \nonumber \\ \left<T(z)W(w)W(v)\right> &=& {\pm 5\sqrt{i}c_{BH}\over 2(z-w)^{2}(w-v)^{4}(v-z)^{2}},  \quad
c=-ic_{BH}
\een
\ben{hr18}
\left<T(z)T(w)T(v)\right> &=& {\pm \left(i\right)^{-{1\over 2}}c_{BH}\over(z-w)^{2}(w-v)^{2}(v-z)^{2}}, \nonumber \\
 \left<T(z)W(w)W(v)\right>&=&{\pm 5\left(i\right)^{-{1\over 2}}c_{BH}\over 2(z-w)^{2}(w-v)^{4}(v-z)^{2}}, \quad
 c = ic_{BH}
\een
The overall sign in the three point functions are not relevant since they can be removed by redefining $(T,W) \to -(T,W)$, without affecting the two point correlators. The three point correlators for the stress tensor $T(z)$ have been evaluated in \cite{Arutyunov:1999nw,Grumiller:2009mw} and are seen to agree with the first choice, a real positive central charge \refb{hr15}. Therefore, for consistency with the asymptotic symmetry algebra \cite{Campoleoni:2010zq}, we should also get the $\left<TWW\right>$ correlators consistent with the first choice in \refb{hr15}. For this we need to know the coupling of the spin-3 field with metric in the Fronsdal formulation. Recently the coupling of spin-3 field with gravity in the metric formulation upto terms quadratic in the spin-3 field has been considered \cite{Campoleoni:2012hp}. The full second order formulation of this theory has been presented in \cite{Fujisawa:2012dk}. This is sufficient to calculate these three-point functions and check the consistency requirement. We hope to return to this in the future. 
\section{Analytic Continuation to dS$_3$-CFT$_2$}\label{ds}
In this section we will consider spin-3 and spin-2 field coupled to background de-Sitter gravity in three dimensions and  we will compute the two point correlators $\left<W_{zzz}W_{zzz}\right>$, $\left<W_{\bar{z}\bar{z}\bar{z}}W_{\bar{z}\bar{z}\bar{z}}\right>$, $\left<T_{zz}T_{zz}\right>$ and $\left<T_{\bar{z}\bar{z}}T_{\bar{z}\bar{z}}\right>$ on $\hat{\II}^{-}$ which is the past infinity of the region $\OO^{-}$ which comprises the causal past of a timelike observer in de-Sitter space. The metric for a planar slicing of $\OO^{-}$ is given as \cite{Strominger:2001pn}
\be{ds1}.
ds^2=\ell^2\left[e^{-t}dzd\bar{z}+dt^2\right].
\ee
By the coordinate transformation $\tau=e^{t}$, we get the metric
\be{ds2}
ds^2=\ell^{2}\left[{-d\tau^2+dzd\bar{z}\over \tau^2}\right].
\ee
The region $\hat{\II}^{-}$ corresponds to $t\to -\infty$ or $\tau=0$. We can, therefore, analytically continue our results for Euclidean AdS$_3$ from the previous sections to dS$_3$ by $\ell\to i\ell$ and $\sigma\to i\tau$ i.e.
\be{ds3}
iS_{ds3}=S_{AdS3}\left[\ell\to i\ell, \sigma\to i\tau\right].
\ee
This analytic continuation leads to the following transformation on the auxilliary variables $U^{A}$, fields $\phi^{(s-r)}$ and the gauge transformation paramters $\epsilon^{(s-r)}$.
\be{ds31}
v\to iv, \quad u^{i}\to u^{i}, \quad \phi^{(s-r)}\to(-i)^{r}\phi^{(s-r)}, \quad \epsilon^{(s-r)}\to(-i)^{r}\epsilon^{(s-r)}.
\ee
By this anaytic continuation the full (Bulk + Boundary) action for spin-3 and spin-2 field in the background \refb{ds2} becomes
\ben{ds4}
S_{ds3}^{(3)} &=&S_{bulk}^{(3)}+S_{bdy}^{(3)}+S_{bdy}^{(CT)}, \nonumber \\
S_{ds3}^{(2)} &=& S_{bulk}^{(2)}+S_{bdy}^{(2)},
\een
where
\ben{ds5}
S_{bulk}^{(3)} &=& {3\ell^3 \over 32\pi G_N}\int_{\tau_B}^{\infty}{d\tau\over \tau^{3}}\langle\langle\Phi^{(3)}\left|\left(1-{1\over4}U^2\p_{U}^{2}\right)\FF^{(3)}\Phi^{(3)}\right.\rangle\rangle, \nonumber \\
S_{bdy}^{(3)} &=&  {3\tau_{B}^{-2}\ell\over 32\pi G_N}\left[\left<\left<\phi^{(3)}_{3}\left|\left(-1+{1\over 2}u^2\p_{u}^{2}\right)\chi^{(3)}_{3}\right.\right>\right>_{\tau_{B}}+{1\over 6}\left<\left<3\p_{u}^{2}\phi^{(2)}_{3}-\phi^{(0)}_{3}\left|\chi^{(0)}_{3}\right.\right>\right>_{\tau_{B}} \right. \nonumber \\
&& +\left. {1\over 2}\left<\left<\p_{u}^{2}\phi^{(3)}_{3}+\phi^{(1)}_{3}\left|\zeta^{(1)}_{3}\right.\right>\right>_{\tau_{B}}+{1\over 18}\left<\left<\zeta^{(1)}_{3}\left|\zeta^{(1)}_{3}\right.\right>\right>_{\tau_{B}}\right], \nonumber \\
S_{bdy}^{(CT)} &=& {3\tau_{B}^{-2}\ell\over 16\pi G_N}\left[\left<\left<\phi^{(3)}_{3}\left|G^{(3)}\right.\right>\right>_{\tau_{B}}+{1\over 18}\left<\left<\left(\phi^{(0)}_{3}-3\p_{u}^{2}\phi^{(2)}_{3}\right)\left|K^{(0)}\right.\right>\right>_{\tau_{B}}\right], \nonumber \\
S_{bulk}^{(2)}&=& {\ell^3 \over 32\pi G_N}\int_{\sigma_B}^{\infty}{d\sigma\over \sigma^{3}}\langle\langle\Phi^{(2)}\left|\left(1-{1\over4}U^2\p_{U}^{2}\right)\FF^{(2)}\Phi^{(2)}\right.\rangle\rangle, \nonumber \\
S_{bdy}^{(2)} &=& {\ell\tau_{B}^{-2}\over 32\pi G_N}\langle\langle\phi_{2}^{(2)}\left|\left(-1+{1\over 2}u^2 \p_{u}^{2}\right)\chi_{2}^{(2)}\right.\rangle\rangle_{\sigma_{B}},
\een
where, the definition for all the terms appearing above changes for de-Sitter space
\ben{ds6}
\FF^{(s)} &\equiv&\left(\p_{U}.\DD\right)\left(U.\DD\right)-2\left(U.\DD\right)\left(\p_{U}.\DD\right)+{1\over 2}\left(U.\DD\right)^{2}\p_{U}^{2}+{s+1\over \ell^2}U^2\p_{U}^{2}-{4s\over l^2}, \nonumber \\
\DD_{A} &\equiv&\bar{e}_{A}^{M}\p_{M}-\bar{e}_{A}^{M}\omega_{MB}^{~~~C}U^{B}\p_{U^C}={\tau\over \ell}\p_{A}+{1\over \ell}\left[U_A \p_{U^{\tau}}-U_{\tau}\p_{U^{A}}\right], \nonumber \\
\chi^{(3)}_{3}&\equiv& \p_{\tau}\left(\tau\phi^{(3)}_{3}\right)-\tau\left(u.\p\right)\phi^{(2)}_{3}+u^2\phi^{(1)}_{3}, \nonumber \\
\chi^{(0)}_{3}&\equiv& {1\over 2}\tau\p_{\tau}\left(-3\p_{u}^{2}\phi^{(2)}_{3}+\phi^{(0)}_{3}\right)+3\phi^{(0)}_{3}, \nonumber \\
\zeta^{(1)}_{3}&\equiv& {1\over 2}\tau u.\p\left(3\p_{u}^{2}\phi^{(2)}_{3}-\phi^{(0)}_{3}\right)-9\phi^{(1)}_{3}, \nonumber \\
G^{(3)}&\equiv& c\left(1-{1\over 4}u^2\p_{u}^{2}\right)A^{(3)}, \nonumber \\
K^{(0)}&\equiv& c{\tau\over 4}\left(\p.\p_{u}\right)\p_{u}^{2}A^{(3)}, \nonumber \\
A^{(3)} &\equiv& -\tau^{2}\FF_{2}\phi^{(3)}_{3}-{1\over 36}\tau^{3}\left(u.\p\right)^{3}\left(\phi^{(0)}_{3}-3\p_{u}^{2}\phi^{(2)}_{3}\right), \nonumber \\
\chi_{2}^{(2)}&\equiv& \sigma\p_{\sigma}\phi_{2}^{(2)}-\sigma\left(u.\p\right)\phi_{2}^{(1)}-{1\over 2}u^2\phi_{2}^{(0)}.
\een
The Fronsdal equation $\FF\Phi=0$ for the spin-3 field has the following gauge invariance
\ben{ds7}
\delta\phi^{(3)}_{3} &=& {1\over \ell}\left[\tau\left(u.\p\right)\epsilon^{(2)}_{3}+u^{2}\epsilon^{(1)}_{3}\right], \nonumber \\
\delta\phi_{3}^{(2)} &=&{1\over\ell}\left[\left(\tau\p_{\tau}+2+u^2\p_{u}^{2}\right)\epsilon^{(2)}_{3}+\tau\left(u.\p\right)\epsilon^{(1)}_{3}\right],\nonumber \\
\delta\phi^{(1)}_{3} &=& {1\over \ell}\left[\tau\left(u.\p\right)\p_{u}^{2}\epsilon^{(2)}_{3}+2\left(\tau\p_{\tau}+1\right)\epsilon^{(1)}_{3}\right],\nonumber \\
\delta\phi^{(0)}_{3} &=& {3\over \ell}\tau\p_{\tau}\p_{u}^{2}\epsilon^{(2)}.
\een
The spin-2 equation of motion has the gauge invariance
\ben{ds71}
\delta\phi_{2}^{(2)} &=&{1\over \ell}\left[\sigma\left(u.\p\right)\epsilon^{(1)}_{2}+u^{2}\epsilon^{(0)}_{2}\right],\nonumber \\
\delta\phi^{(1)}_{2} &=&  {1\over \ell}\left[\left(\tau\p_{\tau}+1\right)\epsilon^{(1)}_{2}+\tau\left(u.\p\right)\epsilon^{(0)}_{2}\right],  \nonumber \\
\delta\phi^{(0)}_{2} &=& {2\over \ell}\tau\p_{\tau}\epsilon^{(0)}_{2}.
\een
The action \refb{ds4}, is invariant under the above gauge transformation, for non-vanishing $\epsilon^{(2)}_{3}$, $\p_{\tau}\epsilon^{(2)}_{3}$, $\p_{\tau}\epsilon^{(1)}_{3}$, $\epsilon^{(1)}_{2}$, $\p_{\tau}\epsilon_{2}^{(1)}$ and $\p_{\tau}\epsilon^{(0)}_{2}$ at the boundary, but has an anomaly for non-vanishing $\epsilon^{(1)}_{3}$ and $\epsilon^{(0)}_{2}$ at the boundary. The anomaly is
\ben{ds8}
\delta_{\epsilon^{(1)}_{3}}\left(S^{(3)}+S_{bdy}^{(3)}+S_{bdy}^{(CT)}\right) &=& {3\over 32\pi G_N}\left<\left<\epsilon^{(1)}\left|\AAA^{(1)}_{3}\right. \right>\right> ,\nonumber \\
\delta_{\epsilon_{2}^{(0)}}\left(S^{(2)}+S_{bdy}^{(2)}\right) &=& -{1\over 32\pi G_N}\langle\epsilon_{2}^{(0)}\left|\AAA_{2}^{(0)}\right.\rangle\rangle,
\een
where,
\ben{ds9}
\AAA^{(1)}_{3} &\equiv& -\tau^{-2}\left[-{1\over 2}+2c+c{\tau^{2}\over 6}u.\p \p_{u}.\p\right] \p_{u}^{2}A^{(3)}, \nonumber \\
\AAA^{(0)}_{2}&=& \left[\p^{2}\p_{u}^{2}-\left(\p_{u}.\p\right)^{2}\right]\phi_{2}^{(2)}.
\een
We will now analytically continue our AdS$_3$ solutions \refb{hr7}, to dS$_3$ by absorbing an extra factor of $-i$ onto $h^{(3)}$
\ben{ds10}
\phi^{(3)}_{3}= \tau^{-1}U_2 \left(iq\tau \right)h^{(3)}_{3}, &&\left| ~\phi_{2}^{(2)}=U_{1}\left(iq\tau\right)h^{(2)}_{2} \right.,\nonumber \\
\phi^{(2)}_{3} = {q^2\over 2}U_{1}\left(iq\tau\right)\bar{\rho}^{(2)}_{3}-U_2\left(iq\tau\right)u.\p \rho^{(1)}_{3}, &&\left| ~ \phi^{(1)}_{2}=q^2\tau K_{0}\left(iq\tau\right)\bar{\rho}^{(1)}_{2}-\tau U_{1}\left(iq\tau\right)\left(u.\p\right)\rho^{(0)}_{2} \right. ,\nonumber \\
\phi^{(1)}_{3} = -q^2\tau U_1\left(iq\tau\right)\rho^{(1)}_{3}, \quad \phi^{(0)}_{3}=0, && \left| ~ \phi^{(0)}_{2}=-2q^2\tau^{2}K_{0}\left(iq\tau\right)\rho^{(0)}_{2}. \right.
\een
The above solutions are pure gauge with gauge parameters\footnote{One can see this by comparing $\phi^{(3)}$ and $\phi^{(2)}$ in \refb{ds10} with the gauge transformations \refb{ds7}, \refb{ds71}, transverse traceless projection \refb{hr4} and the fact that for d=2 $h_{TT}=0$.}.
\ben{ds11}
\epsilon^{(2)}_{3}&=& \ell\tau^{-2}U_{2}(iq\tau)\bar{\rho}^{(2)}[h^{(3)}], \quad
\epsilon^{(1)}_{3} = -\ell\tau^{-1}U_{2}(iq\tau)\rho^{(1)}[h^{(3)}], \nonumber \\
\epsilon^{(1)}_{2} &=& \ell\tau{-1}U_{1}\left(iq\tau\right)\bar{\rho}_{2}^{(1)}, \quad
\epsilon^{(0)}_{2}= -\ell U_{1}\left(iq\tau\right) \rho^{(0)}_{2}.
\een
The on-shell action is determined from $\AAA^{(1)}_{3}$ and $\AAA_{2}^{(0)}$, which from \refb{ds9} is
\ben{ds12}
\AAA^{(1)}_{3} &=&\tau_{B}^{-1}\left[-{1\over 2}+2c\right]U_{2}\left(iq\tau_{B}\right)\p_{u}^{2}\FF_{2} h^{(3)} \tau_{B}\left[-{1\over 2}+2c\right]U_{2}\left(iq\tau_{B}\right)\p^{2}\left(u.\p\right)\left(\p_{u}.\p\right)\rho^{(1)} \nonumber \\
&& +{c\over 6}\tau_{B} U_{2}\left(iq\tau_{B}\right)\left(u.\p\right)\left(\p_{u}.\p\right)\p_{u}^{2}\FF_{2} h^{(3)}+{c\over 6}\tau_{B}^{3}U_{2}\left(iq\tau_{B}\right)\p^{4}\left(u.\p\right)\left(\p_{u}.\p\right)\rho^{(1)}, \nonumber \\
\AAA^{(0)}_{2}&=& U_{1}\left(iq\tau\right)\left[\p^{2}\p_{u}^{2}-\left(\p_{u}.\p\right)^{2}\right]h^{(2)}.
\een
In order to cancel $\OO(\tau_{B}^{-2})$ divergence, we need $c={1\over 4}$. After putting $c={1\over 4}$ in \refb{ds12}, we get the on-shell action as
\be{ds13}
S_{on-shell}^{tot}=S^{(3)}+S^{(2)},
\ee
where,
\ben{ds131}
S^{(3)}&=& {3\over 32\pi G_N}\left<\left<\epsilon^{(1)}\left|\AAA^{(1)}_{3}\right. \right>\right> \nonumber \\
&=& {\ell\over 256\pi G_N}\int d^2 x~ \p^{k}\rho^{(1)}_{k}\left[3\p^{2}\p^{l}h_{l}-2\p^{k}\p^{l}\p^{m}h_{lmk}\right]+\OO(\tau_{B}^{2}) \nonumber \\
&=& {\ell\over 768\pi G_N}\int d^2 x~ h^{ijk}\left({\p_{i}\p_{j}\p_{k}\p_{l}\p_{m}\p_{n}\over \p^{2}}\right)h^{lmn}+local+\OO(\tau_{B}^{2}) \nonumber \\
&=& {\ell\over 48\pi G_N}\int d^2 z\left[h_{zzz}\left({\bar{\p}^{5}\over \p}\right)h_{zzz}+h_{\bar{z}\bar{z}\bar{z}}\left({{\p}^{5}\over \bar{\p}}\right)h_{\bar{z}\bar{z}\bar{z}}\right] +local+\OO(\tau_{B}^{2}) \nonumber \\
&=& -{5\ell\over 2\pi^{2} G_N}\int \int d^2 z~ d^2 w\left[{h_{zzz}(\vec{z})h_{zzz}(\vec{w})\over\left(\bar{z}-\bar{w}\right)^{6}}+{h_{\bar{z}\bar{z}\bar{z}}(\vec{z})h_{\bar{z}\bar{z}\bar{z}}(\vec{w})\over\left(z-w\right)^{6}}\right] +local+\OO(\tau_{B}^{2}) .\nonumber \\
\een
and
\ben{ds132}
S^{(2)}&=& -{1\over 32\pi G_N}\langle\langle\epsilon_{2}^{(0)}\left|\AAA_{2}^{(0)}\right.\rangle\rangle \nonumber \\
&=& {\ell\over 32\pi G_{N}}\int d^2 x~ \rho_{2}^{(0)}U_{1}(iq\tau)U_{1}(iq\tau)\left[\p^{2}h-\p^{l}\p^{m}h_{lm}\right] \nonumber \\
&=& -{\ell\over 64\pi G_N}\int d^2 x ~ h^{ij}\left(\p_{i}\p_{j}\p_{l}\p_{m}\over \p^{2}\right)h^{lm} +local+\OO(\tau_{B}^{2}) \nonumber \\
&=& {-\ell\over 16\pi G_{N}}\int d^2 z~ \left[h_{zz}\left(\bar{\p}^{3}\over \p\right)h_{zz}+h_{\bar{z}\bar{z}}\left({\p}^{3}\over \bar{\p}\right)h_{\bar{z}\bar{z}}\right]+local+\OO(\tau_{B}^{2}) \nonumber\\
&=& {3\ell\over 32\pi^{2} G_N}\int \int d^2 z~ d^2 w\left[{h_{zz}(z)h_{zz}(w)\over\left(\bar{z}-\bar{w}\right)^{4}}+{h_{\bar{z}\bar{z}}(z)h_{\bar{z}\bar{z}}(w)\over\left(z-w\right)^{4}}\right] +local+\OO(\tau_{B}^{2}). \nonumber \\
\een
Our on-shell action is the same as $AdS_3$ (upto a difference in sign for spin-3) with AdS-radius replaced by dS-radius\footnote{Because of the analytic continuation, the anomaly equation changed sign, $\epsilon$ appearing in the anomaly equation changed sign for both spin-3 and spin-2, but while $\AAA$ appearing in the anomaly equation changed sign for spin-3 due to the explicit appearance of $\sigma^{-2}$, it did not change sign for spin-2.}. Because of the change in sign for spin-3, we absorb the negative sign by redefining $\hh^{(3)}=ih^{(3)}$ \footnote{This is just a choice of normalization, but as we discussed in the AdS section, the sign and reality of central charge depends on the three point functions which are computed with the normalization choice of the two point function.} . Thus we get the total on-shell action as
\ben{ds133}
S_{on-shell}^{tot} &=& {5\ell\over 8\pi^{2} G_N}\int \int d^2 z~ d^2 w\left[{\hh_{zzz}(\vec{z})\hh_{zzz}(\vec{w})\over\left(\bar{z}-\bar{w}\right)^{6}}+{\hh_{\bar{z}\bar{z}\bar{z}}(\vec{z})\hh_{\bar{z}\bar{z}\bar{z}}(\vec{w})\over\left(z-w\right)^{6}}\right] \nonumber \\
&& +{3\ell\over 8\pi^{2} G_N}\int \int d^2 z~ d^2 w\left[{h_{zz}(\vec{z})h_{zz}(\vec{w})\over\left(\bar{z}-\bar{w}\right)^{4}}+{h_{\bar{z}\bar{z}}(\vec{z})h_{\bar{z}\bar{z}}(\vec{w})\over\left(z-w\right)^{4}}\right].
\een
We will now try to use the above on-shell action computed in de Sitter background and dS-CFT conjecture to compute two point correlation functions involving stress tensor and spin-3 currents. The dS-CFT programme initiated in \cite{Strominger:2001pn} defines the stress tensor of the boundary CFT through the Noether procedure as 
\be{ds134}
T_{\mu\nu}=-{4\pi\over \sqrt{\gamma}}{\delta S\over \delta \gamma^{\mu\nu}}.
\ee
This is obtained from  the general prescription of Brown and York, motivated by Hamilton Jacobi theory in \cite{Brown:1992br} for a spacetime with boundary, coupled to the constraint that the energy density be real and positive. 
The dS/CFT conjecture of \cite{Maldacena:2002vr} relates the partition function of a CFT to the Hartle-Hawking wavefunction of the universe.
In the semiclassical approximation, it takes the form
\be{ds135}
Z_{CFT}[h]=e^{iS_{cl}[g]}.
\ee
It was also argued (see footnote-23 of \cite{Maldacena:2002vr}), that one may define stress tensor from the CFT patition function with an overall $i$ i.e
\be{ds137}
T_{ij}\equiv i\left({4\pi \over \sqrt{h}}\right)\left({\delta Z[h]\over \delta h^{ij}}\right).
\ee
Semiclassically, this is equivalent to \refb{ds134}. The boundary spin-3 current is the conserved charge associated to the higher-spin symmetry of the bulk, just as the stress tensor is associated to diffeomorphism invariance in the bulk. We will extrapolate the above definition to the spin-3 conserved current in what follows. It would be interesting to prove from first principles that these are indeed the appropriate charges perhaps by extending the AdS proof for stress tensor of \cite{Papadimitriou:2005ii} to the de Sitter case involving both stress tensor as well as higher spin currents.

Higher-point functions may be arrived at by functionally differentiating the one-point function with respect to sources at the boundary. We therefore have
\ben{ds138}
\left<T_{i_1 j_1}(\vec{z_1})T_{i_2 j_2}(\vec{z_1})\cdots T_{i_n j_n}(\vec{z_n})\right>&=&\left(4\pi\over \sqrt{h}\right){\delta\over \delta h^{i_1 j_1}(\vec{z_1})}\left<T_{i_2 j_2}(\vec{z_1})\cdots T_{i_n j_n}(\vec{z_n})\right> \nonumber \\
&=&\cdots= \left(4\pi\over \sqrt{h}\right)^{n}{\delta^{n}\over \delta h^{i_1 j_1}(\vec{z_1})\cdots\delta h^{i^n j^n}(\vec{z_n})}(iZ_{CFT}), \nonumber \\
\left<W_{i_1 j_1 k_1}(\vec{z_1})W_{i_2 j_2 k_2}(\vec{z_2})\cdots W_{i_n j_n k_n}(\vec{z_n})\right>&=&\left(4\pi\over \sqrt{h}\right){\delta\over \delta h^{i_1 j_1 k_1}(\vec{z_1})}\left<W_{i_2 j_2 k_2}(\vec{z_2})\cdots W_{i_n j_n k_n}(\vec{z_n})\right> \nonumber \\
&=&\cdots= \left(4\pi\over \sqrt{h}\right)^{n}{\delta^{n}\over \delta h^{i_1 j_1 k_1}(\vec{z_1})\cdots\delta h^{i^n j^n k_n}(\vec{z_n})}(iZ_{CFT}). \nonumber \\
\een
Semiclassically, this is the same as functionally differentiating the on-shell action (with a negative sign) with fixed boundary condition. We compute the correlators using the above ``differentiate procedure"\footnote{The term is from \cite{Harlow:2011ke} which distinguishes it from the ``extrapolate procedure" to compute correlation functions. To see the difference between the two see \cite{Harlow:2011ke}.}, of \cite{Strominger:2001pn,Maldacena:2002vr}. Since in holomorphic coordinates $\sqrt{h}={i\over 2}$, we have
\ben{ds14}
\left<W_{zzz}(\vec{z})W_{zzz}(\vec{w})\right>&=& (8\pi i)^2\left({\delta^{2}\over \delta \hh^{zzz}(\vec{z})\delta \hh^{zzz}(\vec{w})}\right)\left(-S_{on-shell}^{tot}\right){|_{\hh_{zzz}=\hh_{\bar{z}\bar{z}\bar{z}}=h_{zz}=h_{\bar{z}\bar{z}}=0}} \nonumber \\
&=&{5\ell\over 4G_N \left(z-w\right)^{6}} = {5 c_{ds}\over 6 \left(z-w\right)^{6}}, \nonumber \\
\left<W_{\bar{z}\bar{z}\bar{z}}(\vec{z})W_{\bar{z}\bar{z}\bar{z}}(\vec{w})\right>&=& (8\pi i)^2\left({\delta^{2}\over \delta \hh^{\bar{z}\bar{z}\bar{z}}(\vec{z})\delta \hh^{\bar{z}\bar{z}\bar{z}}(\vec{w})}\right)\left(-S_{on-shell}^{tot}\right){|_{\hh_{zzz}=\hh_{\bar{z}\bar{z}\bar{z}}=h_{zz}=h_{\bar{z}\bar{z}}=0}} \nonumber \\
&=& {5\ell\over 4G_N \left(\bar{z}-\bar{w}\right)^{6}} = {5 c_{ds}\over 6 \left(\bar{z}-\bar{w}\right)^{6}}, \nonumber \\
\left<T_{\bar{z}\bar{z}}(\vec{z})T_{\bar{z}\bar{z}}(\vec{w})\right>&=& (8\pi i)^2\left({\delta^{2}\over \delta h^{\bar{z}\bar{z}}(\vec{z})\delta h^{\bar{z}\bar{z}}(\vec{w})}\right)\left(-S_{on-shell}^{tot}\right){|_{\hh_{zzz}=\hh_{\bar{z}\bar{z}\bar{z}}=h_{zz}=h_{\bar{z}\bar{z}}=0}} \nonumber \\
&=& {3\ell\over 4G_N \left(\bar{z}-\bar{w}\right)^{4}} = {c_{ds}\over 2 \left(\bar{z}-\bar{w}\right)^{4}}, \nonumber \\
\left<T_{zz}(\vec{z})T_{zz}(\vec{w})\right>&=& (8\pi i)^2\left({\delta^{2}\over \delta h^{zz}(z)\delta h^{zz}(w)}\right)\left(-S_{on-shell}^{tot}\right){|_{\hh_{zzz}=\hh_{\bar{z}\bar{z}\bar{z}}=h_{zz}=h_{\bar{z}\bar{z}}=0}} \nonumber \\
&=& {3\ell\over 4G_N\left(z-w\right)^{4}}= {c_{ds}\over 2 \left(z-w\right)^{4}},
\een
where $c_{dS}={3\ell\over 2G_N}$ is the central charge associated with $dS_3$ by analyzing the stress-tensor on $\hat{\II}^{-}$ \cite{Strominger:2001pn}. In order to show that our results are compatible with this central charge, we have to show that the three point functions $\left<TWW\right>$ are given by the first choice in \refb{hr15} with $c_{BH}$ replaced by $c_{dS}$ (which is the same as replacing AdS-radius by dS-radius). We would also like to stress that it is unlikely that our results for two point functions are compatible with the imaginary central charge found in \cite{Ouyang:2011fs}. For this to be true, we should get three point functions $\left<TWW\right>$ according to \refb{hr17} or \refb{hr18}. There is no obvious way that the three point function will pick up a $\sqrt{i}$, either in the numerator or denominator, because the only factors of $i$ will come from $(4\pi i)^3$ or from $h^{(3)}\to i\mathfrak{h}^{(3)}$. We defer any further discussion to after the computation of the three-point function, which we hope to return to in future. 
\section{Conclusions and future directions}
In this paper we used the boundary action for spin-3 coupled to Anti-de Sitter gravity in three dimensions derived in \cite{Joung:2011xb} and employed holographic renormalization to fix the counter term action and compute the two point correlators $\left<WW\right>$ and $\left<TT\right>$. We found that these correlators factorise holomorphically and that our expressions are consistent with the general requirements of 2d conformal invariance. We discussed the consistency requirement of our normalization on the three point functions $\left<TWW\right>$, which we would like to revisit in future work. We then analytically continued our results for AdS$_3$ to compute the correlators in dS$_3$. We also argued consistency requirements on the three point functions for our results to be compatible with the real central charge $c_{dS}={3\ell\over 2G_N}$, associated with $dS_3$ by analyzing the stress-tensor on the past infinity $\hat{\II}^{-}$ of the region $\OO^{-}$ which comprises the causal past of a timelike observer in de-Sitter space \cite{Strominger:2001pn}. We also discussed why it is quite likely that our two point functions are not compatible with the imaginary central charge obtained in \cite{Ouyang:2011fs}. We can make this picture concrete only after the computation of the three point functions $\left<TWW\right>$, which we hope to address in future. Additionally, we would like to refine our techniques used in this paper to be able to compute the correlation functions in topologically massive higher spin gravity \cite{Bagchi:2011vr} in the same lines of topologically massive gravity \cite{Skenderis:2009nt,Grumiller:2009mw}. This will give us concrete evidence for the logarithmic behavior of the boundary CFT as conjectured in \cite{Bagchi:2011vr}. This is work in progress and we hope to report on it soon.
\section*{Acknowledgements} 
We would like to thank Justin David, Eouhin Joung, Rajesh Gopakumar, Rajesh Gupta, Mu-In Park and Marika Taylor for several very helpful discussions. We would especially like to thank Arjun Bagchi and Arunabha Saha for initial collaboration and several very helpful discussions. SL would like to thank the Harish-Chandra Research Institute for support in the form of a Senior Research Fellowship while part of this work was carried out, and more generally the people of India for their generous support to research in theoretical sciences. The work of BS is supported in part by the ERC Advanced Grant no. 246974, “Supersymmetry: a window to non-perturbative physics”.

\end{document}